
%
\input phyzzx
\doublespace
\PHYSREV
\def\sss{\scriptscriptstyle}
\def\A{{\rm\sss A}}
\def\B{{\rm\sss B}}
\def\C{{\rm\sss C}}
\def\D{{\rm\sss D}}
\def\E{{\rm\sss E}}
\def\F{{\rm\sss F}}
\def\H{{\rm\sss H}}
\def\I{{\rm\sss I}}
\def\II{{\rm\sss II}}
\def\III{{\rm\sss III}}
\def\IV{{\rm\sss IV}}
\def\V{{\rm\sss V}}
\def\VI{{\rm\sss VI}}
\def\VII{{\rm\sss VII}}
\def\VIII{{\rm\sss VIII}}
\def\cf{C_\F}
\def\ca{C_\A}
\def\mh{M_\H}
\def\GeV{{\rm GeV}}
\def\TeV{{\rm TeV}}

\def\a{\alpha}
\def\b{\beta}
\def\g{\gamma}
\def\G{\Gamma}
\def\d{\delta}
\def\e{\epsilon}
\def\la{\lambda}
\def\si{\sigma}
\def\mn{{\mu\nu}}
\def\N{{\cal N}}
\def\O{{\cal O}}
\def\cA{{\cal A}}
\def\ra{\rightarrow}
\def\c{\cdot}
\def\r{\right}
\def\l{\left}
\def\o{\over}
\def\MSbar{\overline {\rm MS}}
\def\Bt{\tilde B}

\def\pslash{p\mkern-8mu{/}}
\frontpagetrue
{\baselineskip 14pt
\null
\line{\hfill BNL-xxxxx}
\line{\hfill October, 1993}
\vskip 2.0in

\centerline{\bf QCD CORRECTIONS TO HIGGS BOSON PRODUCTION:}
\vskip .25in
\centerline{\bf NON-LEADING TERMS IN THE HEAVY QUARK LIMIT}

\vskip .75cm
\centerline{
{\bf  S.~Dawson and R.~Kauffman$^*$}}
\foot{This manuscript has been authored
under contract number DE-AC02-76-CH-00016 with the U.S.
Department of Energy.  Accordingly, the U.S. Government
retains a non-exclusive, royalty-free license to publish
or reproduce the published form of this contribution,
or allow others to do
so, for U.S. Government purposes.}
\vskip 0.5cm
\centerline{ {\it Physics Department}}
\centerline{\it Brookhaven National Laboratory, Upton,
NY~~11973}
\vskip 0.5cm

 We compute analytic results for the QCD corrections to Higgs
boson production via gluon fusion in hadronic collisions in the limit in which
the top quark is much heavier than the Higgs boson.
The first non-leading corrections of $\O(\alpha_s^3 \mh^2/m_t^2)$
are given and
numerical results presented for both LHC and SSC energies.
We confirm earlier numerical results showing that the dominant
corrections have the same mass dependence as the Born cross section.
\vskip 0.25cm

\vfill
}
\chapter{INTRODUCTION}

One of the prime motivations for the construction
of high energy hadron colliders is to unravel the mechanism of
electroweak symmetry breaking.
In the standard model of electroweak interactions
there exists a physical scalar boson, called the Higgs boson, whose
interactions generate the non-zero masses of the $W$ and $Z$ gauge
bosons.\Ref\hhg{
For a review of Higgs boson phenomenology, see
J.~Gunion {\it et. al.}, {\it The Higgs Hunter's
Guide} (Addison-Wesley, Menlo Park, 1991); M.~Chanowitz, {\it
Ann. Rev. Nucl. Part. Phys.} {\bf 38} (1988) 323.}
  The couplings of the Higgs boson
are completely specified in the standard model; the only
unknown parameter is its mass.
For a given mass, therefore, it is possible to
predict the properties and production mechanisms of the
standard model Higgs boson  unambiguously.
In this paper, we discuss the two-loop QCD radiative corrections
of $\O(\alpha_s^3)$
to the production of the Higgs boson in hadronic interactions.

\FIG\ggh{Top quark loop contributing to $gg\ra H$.}
A particularly interesting mass region in which to search for the Higgs
boson is the intermediate mass region, $80\lsim \mh\lsim
150$~GeV.
The dominant decay mode for the intermediate mass Higgs boson is
$H\rightarrow b {\overline b}$.  However, the formidable QCD
background to this decay will probably necessitate using rare decay
modes such as $H\rightarrow \gamma \gamma$ to search for the
intermediate mass Higgs boson.  Since the number of events remaining
after cuts to remove backgrounds is small, it is vital to understand
the effects of radiative corrections in this region in order to
determine the viability of the signal.

In the intermediate  mass region, the
primary production mechanism is gluon fusion through a top
quark loop as shown in Fig. \ggh.  For a heavy top quark, $m_t\gsim
150$~GeV, and an intermediate-mass Higgs boson
it makes sense to expand the results in powers of
$r\equiv\mh^2/m_t^2$.
The zeroth-order result in this expansion, \ie, the $m_t\ra\infty$ limit,
is remarkably accurate, giving the amplitude in Fig. \ggh\ to better than
10\% even up to $r=1$.  However, there is no guarantee that the radiative
corrections will have as little dependence on $r$.

The leading corrections for $m_t\ra \infty$
have been computed previously and found to increase the cross section
by about a factor of two. Thus, it is critical to determine the $m_t$
dependence of the radiatively-corrected cross section. In this paper
we take the natural first step :
we compute the first non-leading corrections
of $\O(\a_s^3 r)$ to Higgs boson production in hadronic collisions.
In such a limit, the computation of the two loop QCD radiative corrections
becomes greatly simplified and it is possible to obtain analytic
results.
\REF\dd{S.~Dawson, {\it Nucl. Phys. } {\bf B359} (1991) 283;
A.~ Djouadi, M.~ Spira and P. Zerwas, {\it Phys. Lett.}
{\bf B264} (1991)441.}
\REF\sp{D.~ Graudenz, M.~Spira, P.~Zerwas,
{\it Phys. Rev. Lett.} {\bf 70} (1993) 1372;
M.~Spira, Ph.D Thesis, Aachen, 1993.}\refmark{ \dd,\sp}
Our analytical results (for the region over which they are valid)
confirm the
numerical results of Ref. \sp, valid for arbitrary $\mh/m_t$.

If the top quark is very heavy it will show up indirectly via
its contribution to radiative corrections to various quantities
such as the $W$ and $Z$ masses.  Indeed, a global fit to existing data from
electroweak processes requires $m_t\lsim 180~\GeV$
for the
consistency of the standard model.\Ref\will{
P.~Langacker, M.~Luo, and A.~Mann, {\it Rev. Mod. Phys.}
{\bf 64} (1992) 87 .}  For the intermediate mass Higgs boson,
say $\mh\sim 100$ GeV, the $m_t\gg\mh$ limit may be
a reasonable approximation.  For a heavy Higgs boson,
such as $\mh\sim 1~\TeV$, the $m_t\ra\infty$ limit clearly
is not valid.   In this case, our results can be used to gauge
the sensitivity of the Higgs boson production rate to
new physics, since any new heavy quarks will contribute to Higgs
production from gluon fusion.  For example, a doublet of heavy
quarks which is degenerate in mass would not contribute to
the $\rho$ parameter, but would contribute to the gluon fusion
production of a Higgs boson.

The organization of the paper is as follows.  In Section 2, we review
previous results and describe how the low energy theorems are used to
obtain the $\O(\a_s^3)$ corrections to the process $gg \ra
H$ in the limit in which $m_t\ra \infty$.
Section 3 contains a description of the calculational techniques
required to compute the two loop integrals occuring in the evaluation of the
virtual diagrams for gluon fusion
when the non-leading terms in $\mh/m_t$ are retained.
In Section
4 we present our analytical results: partonic cross sections
for $g g$, $q \overline{q}$ and $q g \ra H+X$ to  $\O(\a_s^3 r)$.
Section 5 contains numerical results for
Higgs boson production in $pp$ interactions
at $\sqrt{S}=15~\TeV$ and $\sqrt{S}=40~\TeV$.
Finally, we present our conclusions in Section 6.  The appendices
contain details pertaining to the evaluation of the two-loop integrals.

\chapter{PREVIOUS RESULTS}

The lowest order amplitude for the gluon fusion of
a Higgs boson arises  at one loop from the triangle diagram of
Fig.\ggh.
The amplitude is sensitive to all of the quarks but since the
coupling of quarks to the Higgs boson is proportional to their mass
the contribution from light quarks is suppressed. Assuming there are
no heavier quarks, the contribution from
the top quark is dominant over the range in $\mh$ currently allowed by
experiment ($\mh > 60$~GeV)\Ref\mori{T.~Mori, {\it
Proc. of the XXVI International Conference on High Energy Physics},
Dallas, Texas, (1992)1321.}.
The contribution to the amplitude from a single heavy quark with mass $m_q$
has been available in the literature for
some time,\Ref\gghtree{F.~Wilczek, {\it Phys. Rev. Lett.}
{\bf 39} (1977) 1304; J.~Ellis, M.~Gaillard, D.~
Nanopoulos,
and C.~Sachrajda, {\it Phys. Lett.} {\bf 83B} (1979) 339;
H.~Georgi, S.~Glashow, M.~Machacek, and
D.~Nanopoulos, {\it Phys. Rev. Lett.} {\bf 40} (1978) 692;
T.~Rizzo, {\it Phys. Rev. } {\bf D22} (1980) 178.}
$$
\cA_0^{\mu \nu}\biggl(
g^\mu_\A(k_1)g^\nu_\B(k_2) \ra H\biggr)=-{\a_s\o 2\pi v}\d_{\A\B}
                             (k_1\c k_2 g^\mn - k_1^\nu k_2^\mu )
                              \tau_q \l[1 +(1-\tau_q)f(\tau_q)\r]\quad,
\eqn\tree
$$
where
$v^2=(\sqrt{2}G_F)^{-1}=(246~\GeV)^2$, $\tau_q\equiv 4 m_q^2/\mh^2$,
and
$$
f(\tau_q)=\cases{
\biggl[\sin^{-1} \biggl(
\sqrt{1\o\tau_q}\biggr)\biggr]^2, &
if $\tau_q\ge 1$,\cr
-{1\o 4} \biggl[
 \log\biggl({\eta_+\o \eta_-}\biggr)-i\pi
\biggr]^2, &  if $\tau_q < 1$,\cr}
\eqn\fdef
$$
with
$$
\eta_{\pm}\equiv 1 \pm\sqrt{1-\tau_q}
\quad .
\eqn\eqtadef
$$
Taking the limit $r=\mh^2/m_t^2 \ll 1$,
$${\cal A}_0^\mn \ra -{\a_s\o 3\pi v}\N \d_{\A\B}
\biggl[1+{7 r \over 120}(1+\e) \biggr] (k_1\c k_2 g^\mn-k_1^\nu k_2^\mu)
\quad .
\eqn\alim$$
Neglecting contributions from light quarks,
this gives the spin and color averaged cross section,
$$\eqalign{
\si_0(gg\ra H)&\ra{\a_s^2\o \pi}
{\mh^2\o 576 v^2}\N^2
{1\o 1-\e}
\biggl[1+{7r\o 60}(1+\e)\biggr]
  \delta (s-\mh^2) \cr
& \equiv \si_0^\e \mh^2 \d(s-\mh^2) \quad, \cr}
\eqn\treeres
$$
with
$$
\N =
\biggl[\biggl({4\pi\o m_t^2}\biggr)^\e
\Gamma(1+\e)\biggr]\quad ,
$$
where we have computed the amplitude and cross section in $n=4-2\e$ dimensions
for later use.

When the momentum transfer to the Higgs boson is small, or
equivalently in the limit $m_t\gg\mh$,
 the cross section to $\O(\a_s^3)$  for
$gg\ra H$ can be obtained from the effective
Lagrangian,\Ref\letref{A.~Vainshtein, M.~Voloshin,
V.~Zakharov, and M.~Shifman, {\it Sov. J. Nucl. Phys.}
{\bf 30} (1979) 711; A.~Vainshtein, V.~Zakharov, and M.~Shifman,
{\it Sov. Phys. Usp.} {\bf 23} (1980) 429;
M.~Voloshin, {\it Sov. J. Nucl. Phys.} {\bf 44} (1986) 478.}
$$
{\cal L}_{\rm eff}=-{1\over 4}\biggl[ 1-
 {2 H\o v} {\b_\F\o g (1+\delta)}\biggr]
G^{\mu \nu }_\A
G_{\mu \nu}^\A \quad ,
\eqn\leff
$$
where $G_{\mu \nu}^A$ is the gluon field strength tensor
and $\b_F$ is the contribution of the  top quark to the
QCD beta function:
$$
{\b_F\o g}={\a_s
\o 6 \pi}\biggl\{ 1+{\a_s\o \pi} {19\o 4}
\biggr\}\quad .
\eqn\betadef
$$
  The $(1+\delta)$ term arises from a subtlety in
the use of the low-energy theorem.
\Ref\brat{E.~Braaten and J. Leveille,
{\it Phys. Rev. }{\bf D22} (1980) 715;
M.~Drees and K. ~Hikasa, {\it Phys. Lett.}
{\bf B240} (1990) 455.}
Since the Higgs coupling to heavy fermions is
$m_t(1+H/v) {\overline t} t$,
 the counterterm for the Higgs Yukawa coupling is
fixed in terms of the renormalization of the fermion mass and wave
function.  In the $\MSbar$
scheme,
\REF\ab{S.~Adler and W. Bardeen, {\it Phys. Rev.} {\bf D4} (1971) 3045.}
 $\delta= 2 \alpha_s/\pi$.\refmark{\brat,\ab}
In the limit $m_t\gg\mh$, the  effective Lagrangian
 of Eq. \leff~ can be used to obtain
the gluonic radiative
corrections of $\O(\alpha_s^3)$
 from a one-loop calculation.\refmark{\dd,\sp}  This
 serves as a valuable check of the complete two-loop
calculation.  An effective Lagrangian approach can also be used to
obtain the ${\cal O}(\alpha_s^3)$ radiative corrections to the
gluon fusion of a pseudoscalar, $g g\rightarrow A^0$.\Ref\ks{
R.~Kauffman and W. Schaffer, BNL-49061, 1993, to be published
in {\it Phys. Rev. } {\bf D}; A.~Djouadi, M.~Spira, and P.~Zerwas,
{\it Phys. Lett.} {\bf B311} (1993) 255.}

We proceed to calculate the terms of
$\O(\a_s^3 r)$.  Unfortunately, the $\O(r)$ corrections to the
effective Lagrangian are not known and to obtain them would
require a two-loop calculation.  It is simpler to perform
the direct calculation;
the necessary techniques are discussed in the next section.

\FIG\two{ Two-loop diagrams contributing to $g g \ra H$.}
\FIG\real{Real diagrams contributing to $g g\ra gH$
and to $q \overline{q}\rightarrow g H$.}
\chapter{CALCULATIONAL TECHNIQUES }
\REF\dk{S.~Dawson and R.~Kauffman, {\it Phys. Rev. } {\bf D47} (1993), 1264.}
The evaluation of the two-loop diagrams
arising in the virtual corrections to $gg \rightarrow H$
is an extension of the techniques
used in the case of $H\ra\g\g$.\refmark{\dk}
\REF\hoog{F.~Hoogeveen, {\it Nucl. Phys.} {\bf B259} (1985) 19.}
\REF\rhopar{
J.~van der Bij and M. Veltman, {\it Nucl. Phys.} {\bf B231} (1984) 205.}
The basic strategy is to expand the loop integrals
in powers of the external momenta over $m_t$ at every stage.\refmark{\hoog}
This technique has been successfully used to compute the $2-$loop
contribution to the $\rho$ parameter from a heavy top quark.\refmark{\rhopar}
The complete set of two-loop diagrams is shown in Fig. \two.
Each graph  gives a result of the form
$$
\cA_i^{\mu \nu}
 =C \l(  a_i g^{\mu \nu}k_1\cdot k_2+b_i k_1^\nu k_2^\mu
            +c_i k_1^\mu k_2^\nu  \r)\quad,
\eqn\abdef
$$
where the incoming gluons have momenta, polarization indices and colors
as in Eq. \tree and
 $$
C\equiv - \N^2\d_{\A\B}{\alpha_s^2\over 2\pi^2 v}
\quad .
\eqn\cdef$$
Gauge invariance requires that
$$ \sum_i a_i = -\sum_i b_i, \eqn\gaugeinv $$
where the sum runs over all the diagrams.
(The $c_i$ terms do not contribute for on-shell gluons.)
In  order
to reduce the number of tensor structures and deal with scalar quantities
only, we compute three contractions
of each diagram : $\cA_i^\mn g_\mn $,
$\cA_i^\mn k_{1\mu}k_{2\nu}$ and $\cA_i^\mn k_{1\nu}k_{2\mu}$.
{}From the contracted amplitudes
the values of $a_i$ and $b_i$ can easily be found
$$
\eqalign{
a_i=&{\cA_i^\mn \over C (n-2) (k_1\cdot k_2)^2}
\Bigl\{ k_1\cdot k_2 g_\mn
- k_{1\nu} k_{2\mu} - k_{1\mu} k_{2\nu}\Bigr\},
\cr
b_i=&{\cA_i^\mn \over C (n-2) (k_1\cdot k_2)^2}
\Bigl\{ (n-1) k_{1\mu}  k_{2\nu} - k_1\cdot k_2 g_\mn
           + k_{1\nu}  k_{2\mu}\Bigr\}. \crr }
\eqn\stufd
$$

The various
two-loop diagrams have either one, two or three gluon propagators. Diagrams
with one gluon propagator (these are the ones previously calculated in
Ref. \dk) can be written such that the gluon propagator contains no
external momenta. For those diagrams with more than one gluon propagator
we Feynman parametrize to combine the massless
gluon  propagators
(top quark propagators are left alone); the loop momenta are then shifted
to move the external momenta into the top-quark propagators.
The gluon propagators for diagrams V-VIII become
$$\eqalign{
{\rm V{-}VI}&: {1\o q^2(q-k_1)^2} \ra
 {1\o (q^{\prime2})^2}\quad
{\rm where}~q^\prime=q-xk_1\quad, \cr
{\rm VII}&: {1\o (q+k_1)^2(q-k_2)^2}\ra
 {1\o (q^{\prime2}+x(1-x)\mh^2)^2}\quad
\cr
& \qquad \qquad \qquad {\rm where}~q^\prime=q+xk_1-(1-x)k_2 , \cr
{\rm VIII}&: {1\o q^2(q+k_1)^2(q-k_2)^2}
\ra {1\o (q^{\prime2}+xy\mh^2)^3}\quad
{\rm where}~q^\prime=q+xk_1 -yk_2\quad,\cr}
\eqn\feynman
$$
where the integrals over the Feynman parameters $x$ and $y$ are implicit.
For diagrams V and VI
all the external momentum can be shifted into the heavy-quark propagators.
For diagrams VII and VIII products of $k_1\c k_2=\mh^2/2$ and Feynman
parameters remain in the denominators.

The denominators arising from the heavy-quark propagators in Fig. 2 can
be expanded in powers of the external momentum, e.g.,
$$
{1\o (q-k_1)^2-m_t^2}={1\o q-m_t^2}\biggl(1+{2 q\c k_1\o
q-m_t^2}+ ...\biggr)\quad .
\eqn\exp$$
To obtain the terms of $\O(\mh^2/m_t^2)$ each denominator
must be expanded up to terms containing two powers each of $k_1$ and
$k_2$ (in diagram VIII the denominators must be expanded
to one further power because the momentum integrals bring in an inverse
power of $\mh^2$).  The Feynman integrals for diagrams V and VI can
be performed at this stage since the expansion brings the Feynman parameters
into the numerator.
The Feynman integrals for VII and VIII must be performed after
the momentum integration but they  involve
 only  polynomials and
logarithms and are easily done.

After contracting the amplitudes  from
the graphs of Fig. 2 as in Eq. \stufd~ and expanding the denominators
all the  contributions have the form
$$
\int {d^n p\o (2\pi)^n}
 \int {d^n q\o (2\pi)^n}
{({\rm powers~of~} p\c k_i,q\c k_i)\times
 ({\rm powers~of~}p^2,~q^2,~p\c q)
\o [(p+q)^2-m^2]^j(p^2-m_t^2)^k (q^2-m_t^2)^l }\quad,
\eqn\ints
$$
where $m^2$ can be zero or a product of Feynman parameters times $\mh^2$.
Using symmetry arguments the powers of $p\c k_i$ and
$q\c k_i$ in the numerators can be written in terms of powers
of $p^2,~q^2,$ and $p\c q$ times
powers of $k_1\c k_2=\mh^2/2$.  The integrals can then be reduced
to the symmetric form\Ref\pas{G.~
Passarino and M.~Veltman, {\it Nucl. Phys.}
{\bf B160} (1979) 151.}
$$
\int {d^n p\o (2\pi)^n} \int {d^n q\o (2\pi)^n}
{1\o  [(p+q)^2-m^2]^j (p^2-m_t^2)^k (q^2-m_t^2)^l }\quad .
\eqn\ints
$$
These integrals are well known in the literature and are discussed in
Appendix A.  The techniques necessary to symmetrize the numerators are
discussed in Appendix B.

\chapter{RESULTS}

To compute the radiative corrections for the inclusive production of
the Higgs boson from gluon fusion, we need both the
real contribution from $ gg \ra gH$ and the virtual corrections to
$gg\ra H$.   We will also need the contributions for
$q g \ra q g H$ and
$q {\overline q}\ra gH$.

\subsection { (a)  Virtual Corrections to $g g \ra H$}

The results of the
 two-loop diagrams contributing to $gg\ra H$ are
detailed in Appendix C.  The sum of the two-loop amplitudes can be
written in terms of the Born amplitude from
Eq.~\alim, (neglecting the irrelevant imaginary part)
$$
Re(\cA_\V^\mn) = {\a_s \o 2\pi} \cA_0^\mn \N
\biggl[ \ca \biggl( -{r^{-\e}\o \e^2}
+{5\o 2} + {2\pi^2\o3} -{19 r\o 180}\biggr)
-{3 \cf\o 2}\biggl(3+{7r\o60\e}-{13 r\o 1080}\biggr)
\biggr]
+\O(r^2) \quad.
\eqn\vamp$$
The color factors for SU(3) are given by $\ca=3$ and $\cf=4/3$.
\REF\gamgam{H.~Zheng and D.~Wu, {\it Phys. Rev.} {\bf D42} (1990) 3760;
A.~Djouadi, M.~Spira, J.~van der Bij, and P.Zerwas, {\it
Phys. Lett.} {\bf B257} (1991) 187; K. Melnikov and O. Yakovlev,
{\it Phys. Lett.} {\bf B312} (1993) 179;
 A.~Djouadi, M.~Spira, and P. Zerwas,
{\it Phys. Lett.} {\bf B311} (1993) 255.}
The term proportional to $\cf$ contributes to $H\ra \g
\g$ and was computed in Refs. \dk, \gamgam.

It is natural to group
the mass counterterms on the
fermion propagator and those at the $H t \overline{t}$ vertex with
the virtual corrections.
The inclusion of the mass counterterms (which have no $\ca$ component)
renders the contribution proportional to $\cf$ finite.
We use the on-mass shell renormalization scheme in which the physical
mass is defined to be the pole in the propagator.  The renormalized
fermion propagator is then:
$$
{i\o\pslash - m_t}\biggl\{ 1+ {\Sigma_R(p)\o\pslash - m_t} + ... \biggr\}
\eqn\renprop
$$
where \foot{  The various factors of $Z_2$ on the fermion propagators and
at the $H{\overline t} t$ vertex and the factors of $Z_1$ at the $g
{\overline t} t$ vertices combine to give the charge renormalization
counterterm given in Section 4c.}
$$
\Sigma_R(p)=\Sigma(p)-\delta m_t-Z_2 (\pslash-m_t)
\eqn\sigr
$$
and
$$
{\delta m_t\o m_t}=\N \cf {\a_s\o \pi}
\biggl({3\o 4 \e}+1\biggr)
\quad .
\eqn\delm
$$
The sum of the mass counterterms is
$$\cA_{\d m}^{\mu \nu}
= \N {\a_s\o2\pi} \cA_0^{\mu \nu} 3 \cf
              \l( 1 + {7r\o120\e} + {49r\o360} \r) + \O(r^2) \quad.
\eqn\act$$

Our result for the virtual amplitude plus mass counterterms,  is then
($r\equiv \mh^2/m_t^2$):
$$
Re (\cA_{\rm\sss V}) = {\a_s \o 2\pi} \cA_0 \N
\biggl[ \ca \biggl( -{r^{-\e}\o \e^2}
+{5\o 2} + {2\pi^2\o3} -{19 r\o 180}\biggr)
-{3 \cf\o 2}\biggl(1-{307 r\o 1080}\biggr)
\biggr]
+\O(r^2)
\eqn\vans
$$
which gives the contribution to the spin and color averaged cross section:
$$\eqalign{
\si_{\rm virt}=&
\si_0^{\e}{\a_s\o \pi} \N
\biggl\{ \ca \biggl( -{r^{-\e}\o \e^2}
+{5\o 2} + {2\pi^2\o3} -{19 r\o 180}\biggr) \cr
& \qquad\qquad\qquad
 -{3 \cf\o 2}\biggl(1-{307 r\o 1080}\biggr)\biggr\}
\delta(1-z)+\O(r^2)\quad, \cr}
\eqn\vsig
$$
where $z=\mh^2/s$ and $\sigma_0^\epsilon$ is defined in Eq. (2.5)
and contains the overall factor $[1+7r(1+\e)/60]$.

There are various  checks we can perform on the contributions to
Eq. \vamp.  The diagrams VII and VIII have imaginary parts and terms
proportional to $\log r$ and $\log^2 r$.
The imaginary parts can be obtained via the Cutkosky rules
and the $\log r$ terms can be related to them since
they both arise from terms containing $\log(-r)$.
The $1/\e$ terms
coming from diagrams VI and VIII are purely infrared  and can computed
by appropriately contracting the legs of the $gggH$
box diagram.
These various checks are described
in Appendix D.

\subsection{ (b)Real Diagrams for $g g \ra g H$}

\REF\ellis{R.Ellis, I.~Hinchliffe, M.~Soldate and J.~Van der Bij,
{\it Nucl. Phys.} {\bf B297} (1988) 221.}
The  matrix element squared for the process
$g g\ra gH$ can be found from the diagrams of
Fig.~\real.\refmark{\ellis}
  The result, with the appropriate spin and color sums
and averages, is
$$
|\overline\cA(gg\ra gH)|^2=8\si_0^\e \ca \a_s
\biggl\{ {s^4+t^4+u^4+\mh^8 \o stu} -{r \mh^2\o 10}\biggr\} + \O(r^2) \quad.
\eqn\azero
$$
To make clear the structure of this result we write it terms of the
$r=0$ result from Ref. \dd:
$$
|\overline\cA|^2={\si_0^\e \o \si_0^\e (r=0)}
|\overline\cA(r=0)|^2 - {8\o10}\si_0^\e\ca\a_s r \mh^2
                         + \O(r^2) \quad.
\eqn\rzero
$$
We see that the bulk of the terms, and all those which yield infinities,
have the same structure as the $r=0$ result and are scaled by the
$m_t$ dependence of the Born cross section.  Clearly, the soft and
collinear singularities will factor as they must.
The form of the extra $\O(r)$ term in Eq. \rzero\ is determined by
Bose symmetry: $\mh^2$ is the only term symmetric in $s,~t$ and $u$ with
the proper dimension. No terms which grow with $s$ and would spoil
the expansion are allowed.

Integrating over the phase space of the final state gluon, we find the
contribution to the cross section:
$$\eqalign{
\si_{\rm real} =&\si_0^{\e}{\ca\a_s \o \pi}\N r^{-\e}
\biggl\{\biggl({1\o \e^2}-{\pi^2\o3}\biggr)
\delta(1-z)  -{1\o\e}z^{1+\e} {\tilde P}_{gg}(z)
 \cr
&\qquad\qquad+ 2\l[{\log(1-z)\o1-z}\r]_+[1+(1-z)^4+z^4]
-{11\o 6} (1-z)^3
\cr &\qquad\qquad
 -{r z(1-z)\o 20} \biggr\} +\O(r^2)\quad ,\cr}
\eqn\realans
$$
where
$$
{\tilde P}_{gg}(z)=
 2 \biggl\{
\biggl( { z\o 1-z}\biggr)_+
+{1-z\o z}+z(1-z)\biggr\}\quad .
\eqn\pt
$$
The plus distribution functions are defined,
$$
\int {f(x)\over (1-x)_+} \equiv \int {f(x)-f(1)\over 1-x}
\quad .
\eqn
\plus
$$

\subsection{(c) Counterterms for $gg \rightarrow H X$}

 In addition to the mass counterterms included in the virtual
diagrams, there are also counterterms due to the
gluon wavefunction and charge renormalization.

For the charge renormalization, we use a modification
of the ${\overline {MS}}$ scheme in which the top quark
decouples as its mass goes to infinity.\Ref\collins{
J.~Collins, F.~Wilczek, {\it Phys. Rev.} {\bf D18} (1978) 242;
W.~Marciano, {\it Phys. Rev. } {\bf D29} (1984) 580.}
  In this scheme, the renormalized
coupling is related to the bare coupling by,
$$\eqalign{
\a_s^R& = \a_s^0\biggl\{
1-{\a_s b_0\o \e}\N
\biggl({m_t^2\o \mu^2}\biggr)^\e
+{\a_s\o 6 \pi\e}
\N
\biggr\}
\cr
&\equiv\biggl(1+2 \delta Z_g\biggr) \a_s^0
\cr}
\eqn\alphadef$$
where $b_0$ is the QCD
 $\b$ function:
$$
b_0={1\o 4\pi} \biggl({11 \ca\o 3}-{2\o 3}n_{lf}\biggr)\quad .
\eqn\betadef
$$
The charge renormalization then gives a contribution
to the cross section,
$$
\si_{\rm ch}=4 \delta Z_g \si_0^\e \d(1-z)
\quad .
\eqn\charge$$

There is also a counterterm due to the wavefunction renormalization
on the external gluon legs:
$$\si_{\rm wf}= 4 \delta Z_{\rm wf}
  \si_0^{\e} \d(1-z) \eqn\sigwf$$
where
in the $\MSbar$ scheme,
$$
\delta Z_{\rm wf} =-\N {\a_s\o \pi}
 {1\o 12 \e }\quad .
\eqn\zwf
$$
\REF\ap{G.~Altarelli and G.~Parisi, {\it Nucl. Phys.} {\bf B126} (1977) 298.}
Finally, there is the Altarelli-Parisi subtraction\refmark{\ap} which
factors out the soft singularity and gives a
contribution,
$$
\si_{\rm\sss AP}=\N\biggl({m_t^2\o \mu^2}\biggr)^\e
{\a_s\o  \pi\e}
z P_{gg}(z)\si_0^{\e}\eqn\apdef$$
where
$P_{gg}$ is the Altarelli-Parisi splitting function,
$$
P_{gg}(z)=\ca{\tilde P}_{gg}(z)+2 \pi b_0\delta(1-z)\quad .
\eqn\split$$

The final result, the physical
 cross section  for $gg\ra H X$ at next-to-leading order
in $\a_s$, is then the sum
$$
{\hat \si}_{\rm\sss TOT}(gg\ra H X)=
\si_0 + \si_{\rm virt} +\si_{\rm real}+
\si_{\rm ch} +\si_{\rm wf}+\si_{\rm\sss AP}
\quad,
\eqn\anstot
$$
where $\si_0 = \si_0^\e\big|_{\e=0}$ (with $\alpha_s$ evaluated
at $\mu$).
It is convenient to write our result as
$$
\eqalign{
{\hat \si}_{\rm\sss TOT}(gg \ra HX)= & \si_0
\biggl\{
\delta(1-z)+{\a_s(\mu)\o \pi}\biggl[
h(z)+{\overline h}(z)\log\biggl({\mh^2\o \mu^2}\biggr)
\cr
&\qquad\qquad \qquad\qquad
+ {34 r\o 135} \delta (1-z) - {3 r \o 20} z (1-z)\biggr]\biggr\}
+\O(r^2) \cr}\eqn\ans$$
the functions $h(z)$ and ${\overline h}(z)$ are the same
as those of Ref. \dd :
$$\eqalign{
h(z)=& \delta(1-z)\biggl( \pi^2+{11\o 2}\biggr)
-{11\o 2} (1-z)^3 \cr &
+2\ca [1+(1-z)^4+z^4] \l[{\log(1-z)\o 1-z}\r]_+
- \ca z {\tilde P}_{gg}(z) \log z \quad,
\cr
\overline{h}(z)=& \ca z {\tilde P}_{gg}(z).   \cr}
\eqn\hdef
$$
It is clear that
the dominant contributions to the result are just a rescaling
of the $r=0$ result by the ubiquitous factor $1+7r/60$.  Note the
cancellation of the $\log(m_t/\mh)$ terms.

\subsection { (d) $q {\overline q} \ra g H$}

The process $q {\overline q} \ra g H$ proceeds by the diagram of
Fig. 3c.
The resulting spin and color- averaged cross section is easily found:
$$
\si(q {\overline q}\ra g H)
 = {\a_s^3\o \pi^2 v^2} {(1-z)^3\o 486} \bigg|I \biggl(
{r\o z}, r\biggr)\bigg|^2 \quad.
\eqn\qqgh$$
The integral $I$, which arises from the triangle diagram of Fig. 1 with one of
the gluon legs taken off-shell,
has been computed by Bergstrom and Hulth\Ref\berg{L.Bergstrom
and G. Hulth, {\it Nucl. Phys. }{\bf B259} (1985) 137.},
$$
I(a,b) = 3\int_0^1 dx \int_0^{1-x}
        { 1- 4xy \o 1 - ax(1-x-y) - b xy }
\eqn\idef$$

If we expand the result in powers of $1/m_t^2$ as we did for the
gluon gluon scattering, we find a result which grows with $s$
(as compared to the $gg\ra gH$ case which did not have any such terms):
$$
\si(q \overline q)\ra g H)\sim \biggl(1+{11 s+7 \mh^2\o
60 m_t^2}\biggr)\quad .
\eqn\qqlim
$$
Since we intend to integrate over parton energies much larger than
$m_t$ we must use the exact result of Eq. \qqgh~ in order to have
sensible high energy behavior.

\subsection{(e) $q g\ra q  H$}

 The matrix-element squared for the subprocess $q g
\ra g H$ can be obtained by crossing from
$q \overline{q}\ra g H$.  The spin and color averaged cross
section can then be found by integrating over the $n$
dimensional phase space.
Factoring the soft singularity, we
find the physical cross section ${\hat \si}$:
$$\eqalign{
{\hat \si}(q g\ra g H)=&{\a_s(\mu)\o \pi }
\biggl\{
\si_0 z P_{gq}(z) \biggl[{1\o 2}\log\biggl({s\o \mu^2}\biggr)+
{1\o 2}+\log(1-z)\biggr] +  {\si_0\o3}(1-z)(3z-7)
\cr
&+ {2\o3}\si_0\big|_{r=0}\int_0^1 d \omega
\biggl({\mid I(\tau  ,r)\mid^2
-\mid I(0,r)\mid^2\o 1-\omega}\biggr)\biggl[
1+\omega^2(1-z)^2\biggr] \biggl\} \cr}
\eqn\ansqg
$$
where $\tau\equiv -s(1-z)(1-\omega)/m_t^2$, and
$$
P_{gq}={4\over 3 z} \biggl[1+(1-z)^2\biggr]\quad .
\eqn\pgq
$$
  As was the case for the $q \overline q$ process, the
expansion of this expression in inverse powers of $m_t$ leads to
terms which grow with $s$.  Instead, we integrate Eq.~\ansqg~ numerically
and give our results in the next section.

\chapter{NUMERICAL RESULTS}

In this section we present numerical results for Higgs production
in $pp$ collisions at the LHC ($\sqrt{S}=15$~TeV) and the SSC
($\sqrt{S}=40$~TeV.)
We use parton distribution functions from Morfin and Tung
\Ref\mt{J.~Morfin and W.~Tung, {\it
Z. Phys. C} {\bf 52} (1991) 13.}:
the  next-to-leading order set S1
translated into the $\MSbar$
prescription and the lowest order set extracted from the same data.
  Unless
otherwise stated we will always use the renormalization scale $\mu=\mh$.

\FIG\lhcsig{Lowest order (dotted and dashed) and radiatively corrected (solid)
cross section for $pp\rightarrow H X $ at the LHC, $\sqrt{S}=15$~TeV.
The curves labelled LO pdf and NLO pdf use the lowest order and
next to leading order parton distribution functions of Morfin
and Tung, respectively.}

\FIG\sscsig{Lowest order (dotted and dashed)
  and radiatively corrected (solid)
cross section for $pp\rightarrow H X $ at the SSC, $\sqrt{S}=40$~TeV.
}
\FIG\lhck{a:~~  Ratio of the  radiatively corrected
cross section (Eq. \ans)  to the Born cross section of Eq. (2.5)
at the LHC, $\sqrt{S}=15$~TeV and the SSC, $\sqrt{S}=40$~TeV,  with
$m_t=200$~GeV.
b:~~  Ratio of the  qg and $\overline{q}g$
cross sections  to the Born cross section (with NLO pdfs and
2 loop $\alpha_s$) of Eq. (2.5)
at the LHC, $\sqrt{S}=15$~TeV and the SSC,$\sqrt{S}=40$~TeV,  with
$m_t=200$~GeV.}

Figures \lhcsig~ and \sscsig~ contain the
radiatively corrected cross sections for Higgs boson production
at the LHC and SSC respectively for $m_t$ equals $150~\GeV$ and $200~\GeV$.
Although these figures include the contribution from $gg$, $q g$,
${\overline q}g$, and $q {\overline q}$ initial states the result is
completely dominated by the $gg$ initial state.
To gauge the effect of the radiative corrections we also plot two
versions of the Born result : the first is a consistent leading order
result, using $\a_s$, parton distributions and the hard cross section
all at leading order, the second is a hybrid result using $\a_s$ and
parton distributions
at next-to-leading order (NLO)
 convoluted with the hard cross section at Born
level.
Both versions are equally correct as they differ at higher
order.  We see from Figures \lhcsig\ and \sscsig\ that the consistent
leading order result is almost 50\% larger than the hybrid result.
Comparing the NLO result with the two leading order results we see that
it is about a factor of 1.5 larger than the consistent leading order
result and about a factor of two larger than the hybrid result.  The
comparison to the hybrid result implies that the $\O(\a_s^3)$ contribution
is over half of the full NLO result.

To emphasize the significance of the radiative corrections
 we have plotted the ratio of the radiatively-corrected cross section
 to the two different Born results in Fig. \lhck a.
This ratio is often called the  $K$ factor.
Our results are in complete agreement with Ref. \sp, where the
$K$-factor is defined using the consistent leading order result,
as in the dotted curve in Fig. \lhck a.
{}From this figure we see that using a next-to-leading order $\a_s$
and next-to-leading order parton distributions does not give an
improved approximation to the full NLO result but rather it is better
to do a consistent leading order calculation.
Figure \lhck b  shows the ratio of the $qg$
cross sections to the hybrid Born result. The $q\overline{q}$ contribution
is everywhere negligible.

\FIG\del{Ratio of the $\O(\alpha_s^3 r)$
radiatively corrected cross sections to
the $\O(\alpha_s^3)$ results with $r=0$ (labelled
$\sigma(m_t\rightarrow \infty$) for each subprocess.}
Numerically, the major contribution to the NLO cross section
can be obtained by taking the results of the low energy theorem
(the $\O(\alpha_s^3)$ result with $r=0$)
and scaling by the factor $(1+7 r/60)$.  To emphasize this we
plot in
Fig. \del~ the ratio of the $\O(\alpha_s^3 r)$ result to
the $\O(\alpha_s^3)$ result with $r=0$ for each subprocess.
(These curves all have the 2-loop $\alpha_s$ and the non-leading
parton distribution functions.)  We see that for the gluon fusion
result, the answer is well approximated by the $m_t\rightarrow \infty$
results of Ref. \dd~ and \sp.  Furthermore, the deviation from the
$m_t\rightarrow \infty$ result is approximated to better that 1\%
accuracy by the $m_t$ dependence of the lowest order cross section.
For the $q \overline{q}$ and $q g$ subprocesses, the $m_t\rightarrow
\infty$ limit is a poor approximation over almost the entire kinematic
region.  This is clearly due to the terms
which grow like $s/m_t^2$ as discussed in Section 4d.

\FIG\mudep{Lowest order with LO pdfs and
1 loop $\alpha_s$ (dotted) and radiatively corrected (solid)
cross section as a function of $\mu$ for $\mh=50$, 100, 200 and 500 GeV
at the SSC, $\sqrt{S}=40$~TeV.}
Fig. \mudep\ shows the dependence of the consistent leading order result
and the next-to-leading order result on the renormalization/factorization
scale $\mu$ for a range of Higgs boson masses.  We see that, contrary to
naive expectations, the radiative corrections do not generally reduce the
dependence of the cross section on $\mu$.  One can only expect the
next-to-leading order result to have less $\mu$ dependence if the
radiative corrections are small.  Since the radiative corrections are
about 100\% and, being higher order in $\a_s$, are more scale
dependent, the dependence on $\mu$ of the NLO result is more severe
than the leading order result.

\REF\sum{R.~P.~Kauffman, {\it Phys. Rev.} {\bf D44} (1991) 1415;
{\it ibid} {\bf D45} (1992) 1512; I.~Hinchlifffe and S.~Novaes, {\it Phys.
Rev.} {\bf D38} (1988) 3475; C.-P.~Yuan, {\it Phys. ~Lett.} {\bf B283}
(1992) 395.}
\chapter{CONCLUSIONS}
We have computed the $\O(\alpha_s^3r)$ contributions to
$pp\rightarrow gH$.  They are dominated by the  gluon fusion contribution
and typically increase the lowest order cross section by  a factor
of between $1.5$ and $2$. The lowest order cross section
is sensitive to whether the 1-loop or 2-loop $\alpha_s$ is used
and which distribution functions are used.

    The dominant numerical corrections to the
gluon fusion contribution can be found from
the $m_t\rightarrow
 \infty$ $\O(\alpha_s^3)$ results of Refs. \dd~
and \sp~ by rescaling the cross section by the factor $(1+7r/60)$.
The smallness of the $\O(\alpha_s^3 r)$ terms demonstrates
the valididty of the $m_t\rightarrow \infty$ limit for the gluon fusion
subprocess.  Indeed, Ref. \sp~ found that the $m_t\rightarrow
\infty$ results were good to within $15\%$ even for $\mh > m_t$.

The determination of the
transverse momentum distribution of the Higgs boson requires
summing the $\log(\mh^2/p_T^2)$ terms which are important at low
$p_T$.\refmark{\sum}  The $\O(\alpha_s^3r)$ terms which we have
calculated can be used to extend this summation beyond the
leading order.

\chapter{ACKNOWLEDGEMENTS}
We are grateful to W.~ Schaffer and
A.~ Stange for many valuable discussions.  We also
thank M.~Spira for providing us with a copy of his thesis.
\APPENDIX{A}{A. Symmetrizations}

The numerators of the integrals are of the form
$$
p\c k_1^j\; p\c k_2^k\; q\c k_1^m\; q\c k_2^i\; \times
({\rm powers\ of}\ p^2,\ q^2,\ {\rm and}\ p\c q)
$$
where $p$ and $q$ are the loop momenta.
Hence all the numerators can be written as contractions of external
momenta with products of up to six loop momenta.
Since the denominators contain no external momenta, structures such
as
$$
p^{\mu}p^{\nu}p^{\rho}p^{\si}
$$
must be symmetric in all their indices.
Therefore, we make the substitution
$$p^{\mu}p^{\nu}p^{\rho}p^{\si} \ra C_2 V_4^{\mu\nu\rho\si} (p^2)^2 $$
where
$$ V_4^{\mu\nu\rho\si}  = g^\mn g^{\rho\si} +  g^{\mu\rho}g^{\nu\si}
                        + g^{\mu\si}g^{\nu\rho}
$$
and
$$ C_2 = {1 \o n(n+2)} $$
where $n=4-2\epsilon $ is the number of dimensions.

Similar arguments can be made for more complicated combinations.
The remaining permutations of four loop momenta are
$$\eqalign{
p^{\mu}p^{\nu}p^{\rho}q^{\si}&\ra C_2 V_4^{\mu\nu\rho\si} p^2\; p\c q,
\cr
p^{\mu}p^{\nu}q^{\rho}q^{\si}&\ra C_{22}
\big\{ (n+2)[p^2 q^2 - (p\c q)^2] g^\mn  g^{\rho\si}\cr
& \qquad+(n\; (p\c q)^2-p^2 q^2)V_4^{\mu\nu\rho\si}
\big\},
\cr}
$$

For products of six loop momenta we define
$$
V_6^{\mu\nu\rho\a\b\g} =  g^\mn V_4^{\rho\a\b\g}
                       +  g^{\mu\rho}V_4^{\nu\a\b\g}
                       +  g^{\mu\a}V_4^{\nu\rho\b\g}
                       +  g^{\mu\b}V_4^{\nu\rho\a\g}
                       +  g^{\mu\g}V_4^{\nu\rho\a\b}
$$
The symmetrizations of six loop momenta are
$$\eqalign{
p^{\mu}p^{\nu}p^{\rho}p^{\a}p^{\b}p^{\g} &\ra C_3 V_6^{\mu\nu\rho\a\b\g}
                                             (p^2)^3 \quad, \cr
p^{\mu}p^{\nu}p^{\rho}p^{\a}p^{\b}q^{\g} &\ra C_3 V_6^{\mu\nu\rho\a\b\g}
                                             (p^2)^2 p\c q \quad, \cr
p^{\mu}p^{\nu}p^{\rho}p^{\si}q^{\a}q^{\b} &\ra  C_{24} p^2
 \big\{
   (n+4)\l[p^2 q^2 -  (p\c q)^2\r]
  g^{\a\b} V_4^{\mu\nu\rho\si}\cr
 & \qquad +[ n\;(p\c q)^2 - p^2 q^2]V_6^{\mu\nu\rho\si\a\b}
 \big\}  \quad, \cr
p^{\mu}p^{\nu}p^{\rho}q^{\a}q^{\b}q^{\g} &\ra  C_{24}
 \big\{
   (n+4)\l[p^2 q^2 p\c q -  (p\c q)^3\r]
(  g^{\a\b} V_4^{\g\mu\nu\rho} +   g^{\a\g} V_4^{\b\mu\nu\rho}
\cr &\qquad +  g^{\b\g} V_4^{\a\mu\nu\rho} )
+ [(n+2)(p\c q)^3 - 3 p^2 q^2 p\c q]V_6^{\mu\nu\rho\a\b\g}\big\} \quad.
\cr}
$$
The various coefficients are
$$\eqalign{
C_{22} &= {1\o n(n-1)(n+2)}\quad, \cr
C_3 &= {1\o n(n+2)(n+4)}\quad, \cr
C_{24} &= {1\o n(n-1)(n+2)(n+4)} \quad. \cr
}$$

\APPENDIX{B}{B. Integrals}

In this appendix we present the results necessary to obtain
the 2-loop integrals used in this paper. We use dimensional
regularization with $n=4-2\epsilon$.  The integrals form two
basic classes:  the first where all denominators are massive
(arising from diagrams VII and VIII) and
the second where there are massless denominators (arising from
diagrams I-VI).

The first class of integrals has the general form:
$$
B_{jkl}\equiv -2(4\pi)^4
 \int {d^n p \o (2 \pi)^n}
 \int {d^n q \o (2 \pi)^n}
{1\o [(p+q)^2-m^2]^j (p^2-M^2)^k (q^2-M^2)^l}\quad .
\eqn\bdef
$$
We will need $B_{jkl}$ with $j=1,2,3$ and
$j+k+l<10$.
\REF\tv{G.~t'Hooft and M.~Veltman, {\it Nucl. Phys.} {\bf B44}
(1972) 189.}
Using integration by parts on the $B_{1kl}$ leads to the following
recursion relation \refmark{\tv,\hoog}
$$
B_{1,j+1,l}=
{1\o 2 j M^2}\biggl[\l(n-1-2j-m^2
   {\partial\o \partial m^2}\r)B_{1,j,l}
+{\partial\o \partial m^2} B_{1,j,l-1}
-{\partial\o \partial m^2} B_{1,j-1,l}\biggr] .
\eqn\recur
$$
The $B_{2kl}$ and $B_{3kl}$ can be obtained from the $B_{1kl}$
by differentiation:
$$
B_{j+1,k,l}={1\o j}{\partial \o \partial m^2} B_{j,k,l}
\quad . $$
Hence we need only calculate $B_{111}$.  This integral has been
given in a power series in $b\equiv m^2/M^2$ in Ref. \hoog.
However, we need to carry the expansion  in $b$ further than in
 this reference.
We find:
$$\eqalign{
B_{111}=& M^2\biggl\{  {b+2\o\epsilon^2}
+{1\o \epsilon}\biggl(6+3 b-2b\log b \biggr)
-b\biggl(2+{b\o 3} +{b^2\o 30}+{b^3\o 210}\biggr)
\log b +b \log^2 b \cr
&
+14 +b\biggl(-1+{8 \o 9}b+{31\o 450}b^2+
{389 \o 44100}b^3\biggr)\biggr\}\quad .\cr}
\eqn\bIII
$$

For most of the diagrams any infinities come from the loop momentum
integrals and so there is no need to keep terms which vanish as
$n\ra 4$ in the integrals.  The exception is diagram VIII which has
infrared divergences arising in the Feynman integrals.  These divergences
come from terms of $\O(1/m^2)$ arising in the $B_{3jk}$ integrals.  Thus,
the terms of $\O(1/m^2)$ in the $B_{3jk}$ need to be calculated to $\O(\e^2)$.
Rather than rederive the entire series of integrals keeping terms to
$\O(\e^2)$ we notice that the $\O(1/m^2)$ terms come from a particular part
of the integral in Eq. \bdef, that in which $p=-q$.  Therefore,
changing variables to $p\ra p-q$ and dropping $p$ except in the denominator
$m^2$ we find
$$
B_{3kl} \ra  -2(4\pi)^4
 \int {d^n p \o (2 \pi)^n}{1\o (p^2-m^2)^3}
 \int {d^n q \o (2 \pi)^n}{1\o (q^2-M^2)^{k+l}}\quad ,
\eqn\oneoverm
$$
where we have kept only the terms of $\O(1/m^2)$.  Eq. \oneoverm\ can also
be verified by brute force calculation.

The integrals which contain massless denominators are $B_{jkl}$ with
one of the masses taken to zero
$$\Bt_{jkl} = B_{jkl}\Big|_{m=0}.
\eqn\btilde$$
\REF\dhl{A.~Denner, W.~Hollik, and B.~Lampe, CERN-TH.6874/93, April 1993;
R.~Scharf, Diploma Thesis, Wurzburg, 1991.}
We will need the result for $j=1,2$.
For the case of the $\Bt_{1jk}$ one may simply take the limit
$m\rightarrow0$
in the expressions for $B_{1jk}$ since the limit is well defined.
However, the $\Bt_{2kl}$ have infrared divergences and cannot be derived
directly from the $B_{2kl}$.  Instead, using integration by parts we
derive another recursion relation
$$
\Bt_{2kl} = \Bt_{2,k+1,l-1} + (n-2k-3) \Bt_{1,k+1,l}
                 - 2 (k+1)M^2 \Bt_{1,k+2,l} \quad.
\eqn\recurtwo
$$
Since $\Bt_{2k0}=0$,  the $\Bt_{2k1}$ depend only on the $\Bt_{1kl}$
and the rest of the  integrals follow.
Alternatively, one may use the explicit formula of Ref. \dhl:
$$\eqalign{
{\Bt}_{jkl}=&{1\o 2} (-1)^{j+k+l}\biggl({4\pi\o M^2}\biggr)
^{2\e}(M^2)^{4-j-k-l}\cr
&\times {\G(k+l+j-n)\G(k+j-n/2)\G(l+j-n/2)\G(n/2-j)\o
        \G(k+l+2j-n)\G(k)\G(l)\G(n/2)}\quad .\cr}
\eqn\btans
$$
We have checked that the two approaches give identical results.

\APPENDIX{C}{C. Virtual Diagrams for $g g \rightarrow H$}

The results of the various diagrams can be separated into two-gauge
invariant sets: contributions proportional to the group factors
$\cf$ and $\ca$.  The $\cf$ terms are proportional to the diagrams
for $H\ra\g\g$ computed in Ref \dk. We present them here for completeness,
written in the form of Eq. \abdef.
$$
\cA_i^{\mu \nu}
 = - \N^2{\a_s^2 \over 2\pi^2v}
\l(  a_i g^{\mu \nu}k_1\cdot k_2+b_i k_1^\nu k_2^\mu \r)\quad .
$$

In this appendix we present only the real part of each amplitude.
Only diagrams VII and VIII have imaginary parts:  these are given in
Appendix D.
Our results for the diagrams contributing to $H\ra\g\g$ are :
$$\eqalign{
a_\I= &\biggl({\cf\over 16}\biggr)
{1\o45}\biggl[ {135\o r}\l(-{2\o\e^2} + {1\o\e}+{37\o18} \r)
 -{177\o\e} +7 -{r\o14}\l({277\o\e}+{12821\o30}\r)\biggr]\quad,  \cr
b_\I=&\biggl({\cf\over 16}\biggr)
{1\o45} \biggl[ {123\o\e}+47+{r\o7}\l({223\o2\e}
    +{2924 \o 15} \r)\biggr]\quad,\cr
a_\II=&\biggl({\cf\over 16}\biggr)
{1\o45} \biggl[ {270 \o r} \l({2\o\e^2} - {1\o\e}- {1\o18}\r)
   -{6\o\e}-{872\o3}-{r\o7} \l({17\o\e} +{6719\o60}\r)\biggr]\quad,  \cr
b_\II=&\biggl({\cf\over 16}\biggr)
{1\o45}\biggl[ {114\o\e}+{388\o3}+{r\o7}\l({71\o\e}+{566\o5}\r)
       \biggr]\quad,\cr
a_\III =&\biggl({2 \cf-\ca\over 32}\biggr)
 {1\o3}\biggl[-{36\o r}+{4d_r\o\e} +{34\o9} + {1151r\o2700} \biggr]\quad,
       \cr
b_\III =&\biggl({2 \cf-\ca\over 32}\biggr)
 {1\o3}\biggl[-{4d_r\o\e}-{2\o9}-{109r\o1350} \biggl]\quad, \cr
a_\IV =&\biggl({\cf\over 16}\biggr)
 {1\o3}\biggl[{36\o r}+{16d_r\o\e}-{226\o9}+{6287r\o2700} \biggr]\quad, \cr
b_\IV =&\biggl({\cf\over 16}\biggr)
 {1\o3}\biggl[-{16d_r\o\e}+{194\o9}-{361r\o135} \biggr]\quad, \cr }
\eqn\abelian$$
where
$$
d_r\equiv 1+{7 r \over 120}\quad .
\eqn\drdef
$$
\vfill
\eject
The remaining, purely non-abelian, diagrams are
$$\eqalign{
a_\V& = \biggl( {\ca\over 12}\biggr)\biggl[
-{1\o r}\l({9\o4\e}+{15\o8}\r)+{3\o\e} - {85\o72}
     +{r\o180}\l({29\o\e}+{2497\o48}\r)\biggr]\quad, \cr
b_\V&=\biggl( {\ca\over 12}\biggr)\biggl[
-{3\o\e} +{10\o 9} -{r\o180}\l({29\o\e} + {6181\o120}\r)
\biggr]\quad, \cr
a_\VI&=\biggl( {\ca\over 12}\biggr)\biggl[
-{1\o r}\l( {9\o 2\e}+{15\o4} \r)  +{113\o 36} + {r\o 8} \l(- {1\o\e}+
{517\o2700} \r)
\biggr]\quad,\cr
b_\VI&=\biggl( {\ca\over 12}\biggr)\biggl[
 -{59\o 18} +{r\o 8} \l({1\o\e} - {197\o 1350}  \r)\biggr]\quad, \cr
a_\VII&=\biggl( {\ca\over 12}\biggr)\biggl[
 {9\o r}\l(-{3\o\e}+{11\o2}\r)
+    {1\o3}
\l( 13+{43 r \o 60}\r)\log r-{164\over 9}-{73 r \over 120}
\biggr]\quad,    \cr
b_\VII&=\biggl( {\ca\over 12}\biggr)\biggl[
{1\o3} \l(1+{r\o15}\r) \log r-{19\o18}-{r\o30}
\biggr]\quad,      \cr
a_\VIII&=\biggl( {\ca\over 12}\biggr)
{1\o6} \biggl[{135\o r}\l({1\o\e}-{7\o6}\r)
     + \l(-{12r^{-\e}\o\e^2}+8\pi^2\r)d_r
- \l(13 +{r\o60}\r)\log r \cr
   &\quad -{30\o\e}+{257\o3}-{r\o20}\l({92\o3\e}+{239\o18}\r) \biggr]\quad, \cr
b_\VIII&= \biggl( {\ca\over 12}\biggr){1\o6} \biggl[
\l({12r^{-\e}\o\e^2}-8\pi^2\r)d_r -\l(1 +{23r\o30}\r)\log r \cr
    &\quad +{30\o\e} -{125\o6} +{r\o15}\l({23\o\e}+{547\o12}\r) \biggr]
\quad. \cr}
\eqn\nonabelian$$
We also include in our definition of the virtual diagrams
the mass counterterms:
$$a_{\rm mass}=-b_{\rm mass} = \cf \biggl[ 1+{7 \o 120}
{r\o \epsilon} +{7r  \o 36}\biggr]
\eqn\amass
$$

Multiplying by the appropriate number of each type of diagram,
we find our result,
$$\eqalign{
a_{\rm\sss TOT} =& 4a_\I + 2 a_\II + 4 a_\III +  2 a_\IV
  +4 a_\V +2 a_\VI + a_\VII + 2 a_\VIII +a_{\rm mass}\cr
=& \ca \biggl[\biggl( -{ r^{-\epsilon} \over
3 \epsilon^2}+ {2 \pi^2\o 9} \biggr)\l[1+{7r\o120}(1+\e)\r]
 +{5\o 6} +{29 r \over 2160}\biggr]-{\cf\o 2} \biggl[ 1-{61 r\o 270}
\biggr]\cr
=&-b_{\rm\sss TOT}\quad .\cr }
$$
Combining with the Born amplitude of Eq. (2.4), this give the result
of Eq. (4.4) to ${\cal O}(r^2)$.
Note that there are only two topologically distict diagrams of type VI,
whereas there are four of type V; this is because reversing the quark line
in diagram VI is equivalent to exchanging the two virtual gluons.
The terms of $\O(1/r)$ vanish
in the sum, as required by gauge invariance.

\APPENDIX{D}{D. Checks on Virtual Diagrams}

There are several checks which can be performed on the diagrams
shown in Figure 2, VII and VIII.  Both the coefficients of the logarithms
and the ${1/\epsilon}$ terms can be simply obtained.
We begin by noting that these diagrams can be written
in terms of the triangle  diagram of Fig. 1 with both gluons off-shell.
Then, for example, diagram VII can be written as,
$$\eqalign{
\cA_{\VII}^\mn =& 4 \pi \alpha_s f_{\A\C\E} f_{\B\D\E}
 (2 g_\mn g_{\la\si}-g_{\mu\si}g_{\nu\la}-g_{\mu\la}g_{\nu\si})
\cr & \times
\int {d^n q\over (2 \pi)^n} {1\over q^2 }
{1\over (k_1+k_2-q)^2} i \G^{\si\la}_{\C\D}(q,k_1+k_2-q)
\cr}
\eqn\asev$$
where $\G^{\si\la}_{\C\D}(q,k_1+k_2-q)$
is the triangle diagram
of Fig. 1  with both gluons  evaluated  off-shell.

The coefficients of the logarithms can be obtained by using
the Cutkosky cutting rules to find the imaginary
parts of the amplitudes, which amounts to replacing the
massless gluon propagators
by $-2 \pi i\d(q^2)$ and
$-2 \pi i\d((q-k_1-k_2)^2)$
and adding  an overall factor of $1/2$.
The $ggH$ $3$-point function is then given by:
$$
\d(q_1^2)\d_(q_2^2)\G^{\si\la}_{\C\D}(q_1, q_2)=
-{\a_s\over 2 \pi v} \d_{\C\D}
{\cal N}
\biggl\{ {1\over 3} (1+{7 r \over 120})
\biggl({\mh^2\over 2} g^{\si\la}
-q_2^{\si}q_1^{\la}\biggr)
 -{r\over 360} q_2^\la q_1^\si\biggr\}
\eqn\gghvert
$$
Substituting Eq. \gghvert~ into Eq. \asev~ and
evaluating the $d^nq$ integral explicitly, we find
$$
Im (\cA_\VII^\mn)={\a_s^2\over 8 \pi v}\d_{\A\B}\N^2
\biggl({\ca\over 3}\biggr)
\biggl[{\mh^2\o 6}  g^{\mu \nu} \biggl(13 +{43r\over 60}\biggr)
+{1\over 3} k_1^\nu k_2^\mu\biggl( 1+{r\over 15}\biggr)\biggr]
\quad .
\eqn\imsev
$$

 The imaginary part of the box diagram, Figure 2, VIII can be found
in an identical manner:
$$\eqalign{
Im (\cA_\VIII^\mn)=&{\a
_s^2\over 16 \pi v}\d_{\A\B}\N^2
\biggl({\ca\over 3}\biggr)
\biggl\{\mh^2 g^{\mu \nu}\biggl[ \biggl(
{2\over \epsilon}- \log r-{13\over 6}\biggr)
\cr &+{r\o 60}  \biggl({7\over \epsilon} -{1\over 60}
-{7\over 2} \log r\biggr)\biggr]\cr
& \qquad\qquad\quad+ k_1^\nu k_2^\mu\biggl[ \biggl(-{4\over\epsilon}
+2\log r-{1\over 3}\biggr)+{r\o 30}  \biggl(
-{7\over \epsilon} +{7\over 2}\log r-{23\over 3}\biggr)\biggr]
\biggr\}
\cr}
\eqn\imsev
$$
The logarithms are then obtained by noting that they always enter
as $\log(-r)=i\pi +\log r$.

In order to extract the ${1\over \e}$ singularities in diagram VII, we note
that they arise from the region where $q^2=0$ and hence may be obtained by
setting $q^2=0$ in the numerator of Eq. \asev.
In this region $\G$ is easily evaluated analytically.
 The remaining  momentum
integral over $q$ is straightforwardly
performed and the correct $1/\e$ terms
obtained.  The singularities from diagram VIII can be found   in
an identical manner.

\refout
\figout